\documentclass[12pt]{iopart}
\usepackage[bookmarks=true,pdftex, unicode=true]{hyperref}
\usepackage{graphicx}
\usepackage{esint}

\def\ket#1{|\,#1\,\rangle}

\def\opone{\leavevmode\hbox{\small1\kern-3.8pt\normalsize1}}

\newcommand{\beq}{\begin{equation}}
\newcommand{\eeq}{\end{equation}}
\newcommand{\ba}{\begin{eqnarray}}
\newcommand{\ea}{\end{eqnarray}}
\newcommand{\bea}{\begin{eqnarray}}
\newcommand{\eea}{\end{eqnarray}}
\newcommand{\bma}{\begin{subequations}}
\newcommand{\ema}{\end{subequations}}
\newcommand{\bwt}{\begin{widetext}}
\newcommand{\ewt}{\end{widetext}}

\begin{document}

\title[Scalable time reversal of Raman echo quantum memory ]{Scalable time
reversal of Raman echo quantum memory and
quantum waveform conversion of light pulse}

\author{E.S. Moiseev$^{1,2}$, S.A. Moiseev$^{1,2,3}$}
\address{$^1$ Kazan  (Volga Region) Federal University, Russia}
\address{$^2$ Institute for Informatics of Tatarstan Academy of Sciences, Kazan, Russia}
\address{$^3$ Kazan Physical-Technical Institute of the Russian Academy of
Sciences, Russia}
\ead{samoi@yandex.ru}
% \keywords{quantum information, optical quantum memory, waveform conversion,
% photon echo, time reversal dynamics }
\begin{abstract}
 We have found a new hidden symmetry of time reversal light-atom interaction in
the photon echo quantum memory with Raman atomic transition. The time-reversed
quantum memory creates generalized conditions for
ideal compression/decompression of time duration of the input light pulses and
its wavelength. Based on a general
analytical approach to this scheme, we have studied the optimal conditions for
the light field compression/decompression in resonant atomic systems
characterized by realistic spectral properties.
  The demonstrated necessary conditions for the effective  quantum conversion of
the light waveform and wavelength are also discussed for various possible
realizations of the quantum memory scheme.
The performed study promises new capabilities for fundamental study of the
light-atom interaction and  deterministic quantum manipulation of the light
field, significant for quantum communication and quantum computing.
\end{abstract}

\maketitle

\section{Introduction}

Time reversal dynamics of light and multiatomic systems  is in the heart of the
long-term investigation related to the background of classical and quantum
thermodynamics  \cite{Zeh1992},  CPT symmetry \cite{Gibson1976}, symmetry of
quantum mechanics and electrodynamics equations \cite{Fushchich1994}.
Discovery of nuclear spin echo \cite{Hahn1950} and its numerous analogies, such
as electron spin \cite{Blume1958} and photon echoes \cite{Abella1964} have
stimulated  comprehensive study of  irreversibility and  new experimental
methods
providing longer  time reversal dynamics in various multi-particle systems.
Great insight has been achieved in experiments on so called \emph{magic-echo}
for dipoler coupled nuclear spins \cite{Rhim1970}.
 Here, the surprising recovery of  the initial state is realized experimentally
in
strongly interacted multi-particle systems  by using impact effective inversion
of spin Hamiltonian including dipole-dipole interaction of nuclear spins.
Such generic Hamiltonian approach has provided a universal method for
realization of time reversal dynamics in the ensemble of nuclear spins.
Similar approaches promise large capabilities for coherent control of resonant
atoms coupled with each other, various electromagnetic fields and with single
photon fields.
The time reversal methods become especially important for   optical quantum
memory (QM) in quantum repeaters of quantum communications \cite{Briegel1998}
and in quantum computing \cite{Kok2007}.

Optical QM based on the multi-atomic systems has been studied over the last
ten years \cite{Hammerer2010,Simon2010,JPhysB2012}.
Considerable opportunities for storage of arbitrary multi-qubit states of light
have been experimentally realized for the photon echo based QMs
\cite{JPhysB2012,Hedges2010,Usmani2010,Hosseini2011,Bonarota2011,Clausen2011,
Saglamyurek2011}.
The original scheme of \emph{photon echo quantum memory} (PEQM)
\cite{Moiseev2001} has been proposed  in the framework of the light-atom
interaction characterized by the perfect time reversal of the input signal field
absorption  in the  photon echo emission.  It has been clearly recognized in
detailed
study \cite{Moiseev2004}, then generalized to nonlinear regime of the light
atom
interaction \cite{Moiseev2004b} and demonstrated \cite{Kraus2006} by observation
of hidden symmetry of the light-atom equations  (referred in this paper as
CRIB-protocol, see also the review \cite{Tittel2010}).
Elaboration of the generic Hamiltonian approach in Schr\"{o}dinger picture to
this model of light-atom interaction \cite{Moiseev2007} has made it possible
to find the quantum state of echo field for arbitrary  input light field.
CRIB-protocol reproduces the basic concepts of Loschmidt echo for time reversal
dynamics of atoms and light as the same kind of generalized time reversal
quantum mirror to its classical counterparts in \cite{Fink1999,Calvo2008}.
QM techniques have been also proposed for delicate manipulation of single photon
states \cite{Fleischhauer2005}.  In particular the possibility of light pulse
\emph{compression/decompression} (c/d) has been proposed in a special CRIB
scheme (c/d-PEQM technique).
Here CRIB protocol is generalized onto the irreversible domain of the
light-atom
interaction \cite{Moiseev2010} and demonstrated in \cite{Hosseini2009}.
Such technique can provide both acceleration of quantum communication by
shortening photon wave packets or resonant interaction with the atomic media
characterized by narrower resonant lines.
At the same time, incomplete time reversal dynamics restricts the capabilities
of this technique deteriorating its quality and compression/decompression
efficiency.
More significant increase of unitary light pulse compression (also generally
referred to as quantum waveform conversion) has been found for three-wave
mixing of
the signal light pulse with additional chirped light fields
\cite{Kielpinski2011,Lavoie2013}.
 These results have  stimulated the study for new capabilities and
improvements of c/d-PEQM technique where the quantum storage could be
supplemented with deterministic manipulation of the quantum light fields.
For these purposes we have focused our attention on the off resonant Raman
scheme of
PEQM
\cite{Moiseev2011,Moiseev2010c,Nunn2008,Ham2008,Gouet2009,Hosseini2009,
Hosseini2011}  which has already achieved the record-high quantum efficiency
\cite{Hosseini2011}.
 Here, we have found more general conditions providing \emph{scalable time
reversal} (STR) dynamics of the PEQM.
 Usefulness of STR-symmetry for the quantum waveform conversion of the light
pulse i.e. for  c/d temporal duration and  deterministic wavelength conversion
is demonstrated for transverse and longitudinal IBs.
We also shown a strategy  for  switching control laser fields providing
high quantum efficiency of the studied QM scheme.

\section{Basic model and equations}
Detailed study of spectral properties, mis-phasematching, absorption and
dispersion effects in CRIB-protocol have been performed  in
\cite{Moiseev2004}  for atomic ensembles with $\Lambda$-scheme of quantum
transitions characterized by Doppler \emph{inhomogeneous broadening} (IB).
Here, the rephasing of excited atomic coherence  is realized via
inversion of the atomic detunings on the resonant transition (due to the
change of the Doppler shift) and irradiation of the echo field in the backward
direction to the signal field propagation.
The CRIB-protocol has been extended to the gaseous atomic ensembles with
$\Xi$-atomic transition \cite{Moiseev2004c} and to solid state systems with
microwave resonant line \cite{Moiseev2003} where IB inversion can be realized
by changing the sign of the dipole-dipole interactions with surrounding
controlling spins.
Usage of external electric fields for inversion of IB has been proposed in
\cite{Nilson2005,Kraus2006,Alexanderprl2006} for solid state media.
Authors \cite{Alexanderprl2006} have experimentally demonstrated such kind of
IB inversion on rare-earth ions in inorganic crystal by switching the polarity
of external gradient electric field.
This technique has been further extended for magnetic field gradient  in
\cite{Hetet2008,Hosseini2012}.
Such spectral control of resonant line characterized by spatially
distributed of IB (longitudinal broadening) has opened up new valuable
opportunities for PEQM since an  efficient echo field irradiation becomes
possible in the forward geometry.
 The technique known now as a gradient echo memory (GEM) simplifies
experimental realization of PEQM and extends CRIB-protocol to incomplete time
reversal light-atom dynamics.
Finally GEM scheme can be realized for the Raman atomic transitions as well as
\cite{Hosseini2012}.
However, any loss of perfect time-reversibility could decrease the
fidelity of the input signal retrieval in GEM-protocol even for large quantum
efficiency. This is a result of additional phase modulation of irradiated echo
field \cite{Moiseev2008}.
Here, it is worth noting other new elaborated schemes of PEQMs providing
atomic rephasing without using all tools inherent in CRIB-protocol, such as
atomic frequency comb \cite{Riedmatten2008} and the approaches exploiting even
natural IB of resonant transitions  \cite{Moiseev2011b,Damon2011,Ham2012}.
Nevertheless, it was found \cite{Moiseev2012} that all these new protocols can
also suffer from the loss of perfect time-reversibility by decreased quantum
efficiency and fidelity for relatively broadband input light fields.
Thus the observation \cite{Moiseev2012} indicates an important role of the time
reversal symmetry on the possible new generalization of CRIB-protocol. In
particular it  is related to  realization of deterministic quantum
manipulations of quantum light fields.

In this work, we generalized CRIB-protocol without principal loss of
prefect time-reversibility by using off resonant Raman atomic transitions. It
has been shown that the proposed PEQM scheme can work with usual
(transverse),
longitudinal and with more general type of IB.
\begin{figure}
\includegraphics[width=0.6\textwidth,height=0.5\textwidth]{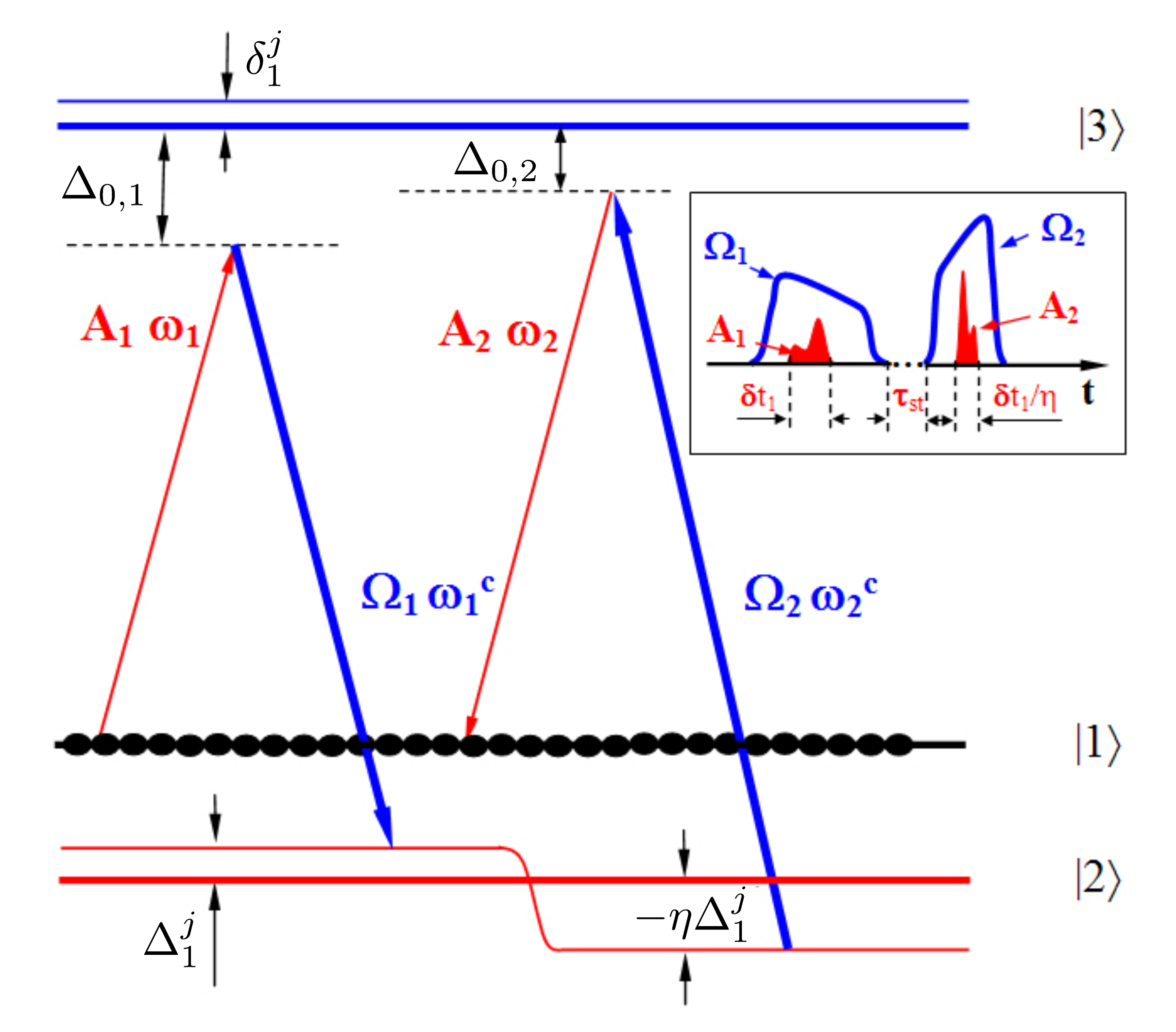}
  \caption{Energy level diagram with atomic transitions for signal/echo pulses
$A_{1,2}$ with carrier frequencies $\omega_{1,2}$ and temporal durations $\delta
t_1$, $\delta t_1/\eta$; control (writing/reading) laser fields have carrier
frequencies $\omega_{1,2}^c $ and Rabi frequencies $\Omega_{1,2}$. Depicted
temporal diagram shows scaled shapes and temporal duration of the light field
envelopes. }
 \label{RamFig1}
\end{figure}
Energy and temporal diagrams of the interaction scheme are depicted in
Fig.\ref{RamFig1}.

At time $t=0$, the signal light pulse  with electric field component
$\hat{E}_{1}(t/\delta t_1,z)=\hat{A}_{1}(t/\delta t_1,z)\exp \left(i
k_{1}z\right)$ enters into the atomic medium (where $\delta t_1$ is a
temporal duration).
The light pulse is also characterized by the quantum state $|\psi_f\rangle$,
spectral width $\delta\omega_1 \approx {\delta t_1}^{-1}$, carrier frequency
$\omega_1$ and wave vector $k_1\uparrow\uparrow z$.
The atomic ensemble is prepared on the long-lived ground states
$|1_a\rangle=\prod_{j=1}^{N} \ket{1}_{j}$ that provides perfect phasematching
for the backward echo pulse emission.
So the initial state of light and atoms is
$|\Psi\rangle=|\psi_f\rangle|1_a\rangle$.
The atoms are simultaneously exposed to an intense control (writing) laser field
(characterized by index $\mu=1$) propagating along the wavevector $\vec{K}_{1}$
with carrier frequency $\omega_1^c$ and Rabi frequency  $\tilde{\Omega} _{\mu}
(t,\vec{r})=\Omega _{\mu} (t )$ $\exp \left( - i(\omega _{\mu}^c t-\vec
{K}_{\mu} \vec {r})\right)$ on $2\leftrightarrow3$ atomic transition.
Indexes $\mu=1$ and $\mu=2$  indicate stages of the quantum storage and
retrieval of the input light field.
The writing laser field is reduced to zero after complete absorption of
the input light field.
The signal and writing fields are assumed to be in Raman resonance
$\omega_{1}-\omega_{1}^c\approx\omega_{21}$ with sufficiently large optical
spectral detuning $\Delta_{0,1}=\omega_{31}-\omega_{1}$ from the
$1\leftrightarrow 3$ transition. Such off resonant interaction avoids
long-lived excitation of the optical level 3.
Total atom-light Hamiltonian is:
\begin{eqnarray}
\hat{H}= \hat{H}_a + \hat{H}_f + \hat{V}_{a-f}+\hat {V}_c,
\end{eqnarray}
where  $\hat{H}_a  =  \sum^{N}_{j=1} \sum^{3}_{n=1} E^{j}_{n} \hat{P}^{j}_{nn}
\label{atomH} $
is an atomic ensemble Hamiltonian, $\hat{P}^{j}_{nm}=|n\rangle_{jj}\langle m|$,
$E^{j}_{n}$ is an energy of $n$-th level for $j$-th atom,  $\hat{H}_f  = \hbar
\int^{+\infty}_{-\infty}dk \omega_k   \hat{a_k}^{+}\hat{a_k}$ is a Hamiltonian
of weak input signal and retrieved echo light fields. Hamiltonians of the
interaction between
the atoms and quantum light fields $\hat {V}_{a - f}$ and classical laser fields
 $\hat {V}_{c}$ (characterized by its Rabi frequencies $\tilde{\Omega} _{\mu}
(t,\vec{r})$) are

\begin{eqnarray}
\hat {V}_{a - f} & = & - \hbar \sum\limits_{j = 1}^N \sum\limits_{{\mu} = 1}^2
\left\{g \hat {A}_{\mu} (z_j )
\exp\left(i k_{\mu} z_{j}\right)
\hat {P}_{31}^j + h.c. \right\},\nonumber\\
\hat {V}_c & = & - \hbar \sum\limits_{j = 1}^N \sum\limits_{{\mu} = 1}^2
\left\{\Omega _{\mu} (t)\exp \left( - i(\omega _{\mu}^c t - \vec{K}_{\mu}
\vec{r}_j)\right) \hat {P}_{32}^j + h.c.\right\}.
\end{eqnarray}
\noindent
$\left[{A}_{\mu '} (z'),{A}_{\mu}^ + (z)\right]=\delta_{\mu ',\mu}\delta
(z'-z),$ $k_{\mu}=(-1)^{\mu+1}$ $\frac{\omega_{\mu} n_{\mu}}{c}$ and $n_\mu$ are
the wave vector of photons and refractive indexes  in the absence of interaction
with atoms, and $g$ is the photon-atom coupling constant \cite{Scully1997}.

We assume the optical transition $1\leftrightarrow 3 $  has a natural IB for the
atomic frequency detunings $\Delta_{2,j}=\omega_{31,j}-\omega_{31,0}$ (where
$\omega_{m1,j}$ is a resonant frequency of j-th atom and $\omega_{m1,0}$ is a
central frequency of the transition $1\leftrightarrow m $, $m=2,3$ ).
In accordance with CRIB-protocol, we suggest that the atomic frequency detunings
 $\Delta_{1,j}=\omega_{21,j}-\omega_{21,0}$ on the  transition $1\leftrightarrow
2 $  can be controlled on demand.
Here, one can assume various experimental methods for controlling of
$\Delta_{1,j}$,
in particular based on the polarity switching of the external electric/magnetic
(e/m) field gradients \cite{Alexanderprl2006,Hetet2008}.

We write a complete system of Heisenberg equations  for the weak  light field
and atomic operators \cite{Scully1997} where the population of excited optical
atomic levels $2$ and $3$ can be ignored $\hat{P}^{j}_{11}|\Psi\rangle \approx
|\Psi\rangle$,
$\hat{P}^{j}_{22}|\Psi\rangle\approx\hat{P}^{j}_{33}|\Psi\rangle\approx0$ as it
is for other ensemble based QMs.
The system of light-atoms equations for absorption ($\mu=1$) and retrieval ($\mu=2$) stages are simplified by usual transfer to slowly variable atomic coherences $\hat {R}_{12,\mu}^j (t)$ and light field amplitudes $\hat{A}_{\mu,o}(z,t)$. By taking into account the carrier frequencies of the probe (echo) light fields and writing (reading) laser fields, we have 
$\hat {R}_{12,\mu}^j (t)=\hat{P}_{12}^j(t)$ $\exp\left(-i\varphi
_{\mu}(\vec{r},z_j)+i(\omega_{\mu}-\omega_{\mu}^c)(t+(-1)^{\mu}n_{\mu}
z_j/c)\right)$, $\hat {R}_{13,{\mu}}^j (t)=\hat {P}_{13}^j (t) $ $\exp \left(
i\omega _{\mu} (t + ( -1)^{\mu}n_{\mu} z_j / c)\right)$ where $\varphi _{\mu}
(\vec{r},z_j) = -((-1)^{\mu}n_{\mu}\omega _{\mu}^c z_j/c + \vec {K}_{\mu} \vec
{r})$  and for the field operator
$\hat{A}_{\mu,o}(z,t)=\hat{A}_{\mu}(z,t)\exp\left(i\omega_{\mu}t\right)$ of the
signal light pulse envelope.
In the moving system of coordinates: $\tau_1 = t - z/v_{g}$ and $Z=z$  we get:
\begin{eqnarray}
\frac{\partial \hat{R}_{13,1}}{\partial \tau_1} & = & -i (\Delta_{0,1} +
\delta_{1}) \hat{R}_{13,1} +
 i g  \hat{A}_{1,o} + i \Omega_1(\tau_1) \hat{R}_{12,1}
\label{eqn2AbsorbSystem}, \\
\frac{\partial \hat{R}_{12,1} } {\partial \tau_1} & = & -i \Delta_{1}
\hat{R}_{12,1} + i \Omega_1(\tau_1) \hat{R}_{13,1} \label{eqn1AbsorbSystem}, \\
\frac{\partial \hat{A}_{1,o} }{\partial Z} & = & i \frac{\pi n_o g S}{v_g} \iint
\limits^{\infty}_{-\infty} d \delta_1 d \Delta_1 G(\delta_1, \Delta_1;Z)
\hat{R}_{13,1} \label{eqn3AbsorbSystem},
\end{eqnarray}
where $v_{g}$ is a group velocity of signal light in the host medium without
dopant resonant atoms;
the atomic operators are parameterized by its continuous spatial
coordinate $Z$ and  spectral detunings $\delta_{1}$,  $\Delta_{1}$ on the
atomic transitions $1\leftrightarrow3$ and $1\leftrightarrow2$.
The atomic detunings are characterized by normalized spectral distribution
function   $G(\delta_1, \Delta_1;Z) $ of the IBs,  so  the atomic index $j$
will be dropped for convenience;
$n_o$ and $S$ are the atomic density and cross section of the signal (echo)
light fields.
For simplicity we assume also that the atomic detunings $\delta_1$ and
$\Delta_1$ are  uncorrelated for each atom.
This assumption will not make a significant effect for large optical detuning
$\Delta_{0,1}$.

We take interest in two special cases.
In the first  case, we assume a spatially homogeneous distribution of the atomic
parameters
$G(\delta_1, \Delta_1;Z)=G_1(\delta_1)G_2(\Delta_1) $ (called also by
"transverse" IB).
In the second case (GEM \cite{Hosseini2012}), we assume that IB on the atomic
transition $1\leftrightarrow2$ is realized due to spatial (linear or slightly
linear) dependence of the atomic detunings $\Delta_1$ along $Z$-axis of the
medium: $G(\delta_1, \Delta_1;Z) = G_{1} (\delta_1 ) G_2(\Delta_1;Z)$. Here, we
use  $G_2(\Delta_1;Z)=\delta (\Delta_{1} -  \chi_{1} Z)$ for GEM scheme
($\chi_\mu$ is determined by external field gradient due to  Stark (or Zeeman)
effect for storage and retrieval stages ($\mu=1,2$); $\delta_{\mu.in}$ and
$\Delta_{\mu.in}$ are the bandwidths of two IBs for the both stages,
respectively. Below we use notation $G_2(\Delta_1;Z)$ for general case of IB
which coincide with each of two mentioned IBs.

\section{Ideal scalable time-reversibility}

At first we study the case off resonant Raman interaction in Eqs.
(\ref{eqn2AbsorbSystem}),(\ref{eqn1AbsorbSystem}) characterized by very large
optical spectral detuning
 $|\Delta_{0,\mu}|\gg |\Delta_{1,\mu}|$ and
$|\Delta_{0,\mu}|\gg \delta\omega_{\mu}$ (where $\delta\omega_{\mu}$ are the
spectral widths of signal and echo fields).
Here, by taking into account the initial
state (given at $\tau_1=-\infty$) of light and atoms  for
Eq.(\ref{eqn2AbsorbSystem}),
we get for the excited optical coherence
\begin{equation}
\hat{R}_{13,1}(\tau_1, Z) \ket{\Psi} \cong
\frac{1}{\Delta_{0,1}}\left\{g \hat{A}_{1,o}(\tau_1,Z) + \Omega_1(\tau_1)
\hat{R}_{12,1}(\tau_1,Z)\right\}\ket{\Psi},
\label{largedetunr_13}
\end{equation}

\noindent
where $\hat{R}_{13}(-\infty,Z)|\Psi\rangle=0$ was used.

Eq. \ref{largedetunr_13} demonstrates purely adiabatical disappearance of the
optical excitation simultaneously with disappearance of the signal and control
light fields.
By using Eq. (\ref{largedetunr_13}) in
Eqs.(\ref{eqn1AbsorbSystem}),(\ref{eqn3AbsorbSystem}) we get:

\begin{eqnarray}
\frac{\partial \hat{R}_{12,1} } {\partial \tau_1}
& = & -i \left(\Delta_1-\frac{\Omega_1^2 (\tau_1)}{\Delta_{0,1}}\right)
\hat{R}_{12,1} +
i \frac{\Omega_1(\tau_1) }{\Delta_{0,1}} g \hat{A}_{1,o}
\label{eqn1AbsorbSystem2}, \\
\frac{\partial \hat{A}_{1,o} }{\partial Z} & = & i \frac{\beta}{2
\Delta_{0,1}}\hat{A}_{1,o} + i \frac{\beta\Omega_1(\tau_1)}{2g
\Delta_{0,1}}\int\limits^{\infty}_{-\infty}  d \Delta_1 G_2(\Delta_1;Z)
\hat{R}_{12,1} \label{eqn2AbsorbSystem2},
\end{eqnarray}
\noindent
where $\beta=2\pi n_o g^2 S / v_g$.

Eqs.(\ref{eqn1AbsorbSystem2}),(\ref{eqn2AbsorbSystem2}) describe off resonant
Raman interaction of the weak signal light field with three-level atomic medium.
%in the case of large enough IB bandwidth in comparison with the spectral width
%of signal field $\Delta_{1,in}\gg\delta\omega_1$.
We focus our attention on the specific property of Eqs.
(\ref{eqn1AbsorbSystem2}) and (\ref{eqn2AbsorbSystem2}) demonstrating possible
tuning of the efficient absorption coefficient $\alpha_{1}=\frac{\beta
\Omega_1^2 }{\Delta_{1,in} \Delta_{0,1}^2}$.
It can be realized by changing the optical spectral detuning $\Delta_{0,1}$,
bandwidth $\Delta_{1,in}$ of the Raman transition and by changing the Rabi
frequency $ \Omega_1$.
We show that controllable changing of these parameters can provide an additional
scalable time-reversible (STR) symmetry of Eqs. (\ref{eqn1AbsorbSystem2}),
(\ref{eqn2AbsorbSystem2}).
By taking into account new spectral atomic parameters and echo signal
irradiation in the backward direction, we write the light-atom equations for the
both stages - absorption of signal pulse ($\mu=1$) and subsequent echo pulse
irradiation ($\mu=2$):

\begin{eqnarray}
\frac{\partial \hat{M}_{12,\mu}} {\partial \tau_{\mu}}
& = & -i \tilde\Delta_{\mu}\hat{M}_{12,o}^{\mu} + i
\frac{\Omega_{\mu}(\tau_{\mu})}{\Delta_{0,\mu}} g \hat{E}_{\mu,o}
\label{eqn3AbsorbSystem2}, \\
\frac{\partial \hat{E}_{\mu,o} }{\partial Z}
& = &  (-1)^{\mu+1} i \frac{\beta\Omega_{\mu}(\tau_{\mu})}{2g
\Delta_{0,\mu}}\int\limits^{\infty}_{-\infty}  d \tilde\Delta_{\mu}
G_{2,\mu}(\tilde\Delta_{\mu};Z) \hat{M}_{12,\mu} ,
\label{eqn4AbsorbSystem2}
\end{eqnarray}

\noindent
where $\tau_2 = t+ z/v_{gr}$ and $Z=z$ occur for new moving coordinate system of
the echo irradiation, $\hat{E}_{\mu,o}=\hat{A}_{\mu,o} \exp(-i(\beta
Z/2\Delta_{0,\mu}) )$ and $\hat{M}_{12,\mu}=\hat{R}_{12,\mu} \exp(-i(\beta
Z/2\Delta_{0,\mu}) )$.
Also we have assumed slowly varied control fields $\Omega_{\mu}(\tau_{\mu})$ and
appropriate Stark shifts of the Raman resonances for both processes
$\Delta_{\mu}=\tilde\Delta_{\mu}+\frac{\Omega_{\mu}^2}{\Delta_{0,{\mu}}}$,
where $\Delta_2$ is a spectral detuning on the atomic transition $1 \rightarrow
2$ for the echo emission stage  ( i.e. signal and echo fields evolve in the
medium in a presence of constant control laser fields).

\subsection{Transverse broadening}

 In usual cases, the shape of IB $G_{2,\mu}(\Delta_{\mu})$ is characterized by
smooth Gaussian or Lorentzian profile.
It is worth noting additional possibilities for realization of the controlled IB
besides those  discussed in the introduction.
The controlled IBs can be  realized in solid state media for atoms (molecules)
characterized by stochastically oriented frozen permanent dipole moments
\cite{Kraus2006}.
Here, the external electric fields can shift the resonant atomic (molecular)
frequencies due to the interaction with the permanent dipole moments.
The induced Stark shifts will be different for the atoms (molecules)
depending on spatial orientations of its permanent dipole moments.
While, changing the polarity of external electric  field at $\tau= \tilde\tau$
will invert the atomic (molecular) spectral shifts providing the  controlled
inversion of IB.  $\tilde\Delta_{2}(\tau\rangle \tilde\tau)
\rightarrow-\tilde\Delta_1$.
Such spectral inversion  of IB will recover the excited atomic coherence leading
to the echo signal emission, respectively.

Also one can use the atomic ensembles in optical waveguides
\cite{Saglamyurek2011} or on the interface of two media with different
refractive indexes \cite{Moiseev2010c}.
Here, the controlled IBs can be a result of spatially inhomogeneous optical
Stark shifts depending on the atomic coordinate in the waveguide crossection or
on the atomic distance from the interface.
While the opposite Stark shifts can be induced by changing the carrier
frequencies of controlled intensive laser field.

Time-reversibility of Eqs. (\ref{eqn3AbsorbSystem2}), (\ref{eqn4AbsorbSystem2})
is revealed by the following symmetry transformation \cite{Kraus2006} for the
standard CRIB-protocol:

\textit{i)} $\tilde\Delta_2\rightarrow-\tilde\Delta_1$,
\textit{ii)} $Z \rightarrow -z$,
\textit{iii)} $\tau_2\rightarrow -\tau_1$,
\textit{iv)} $\hat{E_{2,0}}\rightarrow - \hat{E_{1,o}}$.

\noindent
where the conditions $\textit{i)}-\textit{iv)}$ are met for the light-atoms
equations Eqs. (\ref{eqn3AbsorbSystem2}), (\ref{eqn4AbsorbSystem2}) together
with \emph{coupling} condition
$\Omega_{2}/\Delta_{0,2}=\Omega_{1}/\Delta_{0,1}$.
 An opposite sign of the echo field $\hat{E}_2$  reflects  probably the
well-known difference of the relative phases between the light field and
excited atomic dipole for the absorption or for the emission processes.
This becomes clear if we take into account that  Eqs.
(\ref{eqn3AbsorbSystem2}),
(\ref{eqn4AbsorbSystem2}) are held for similar transformation but where the
condition $\textit{iv)}$ is replaced by $\textit{iv)}^{\prime}:
\hat{M}_{12,2}\rightarrow - \hat{M}_{12,1}$.
Inversion of the excited atomic dipole moments on the resonant transition ($\ket{1} \leftrightarrow \ket{2}$) can be performed by additional  $2\pi$-
pulse on the adjacent atomic transition ($\ket{2} \leftrightarrow \ket{3}$). This  well-known property of $4\pi$ symmetry of two-level system has been realized experimentally in the QED cavity scheme
\cite{Moriti2002}.

Preserving the \emph{coupling} condition  by appropriate variation of the
detuning $\Delta_{0,2}$ and Rabi frequency $\Omega_2$ provides the unitary
wavelength conversion \cite{Moiseev2011}.
 However, we can break the \emph{coupling condition} transfering to more
general symmetry transformation of Eqs.(\ref{eqn3AbsorbSystem2}),
(\ref{eqn4AbsorbSystem2}) that makes possible a scalable time reversibility
(STR) of echo field retrieval.
In turn STR will provide an ideal c/d of the input light temporal duration.
In order to find the STR conditions we note that the ideal quantum compression
of the light pulse duration $\delta t_{e}=\delta t_{1}/\eta$  is
accompanied  by appropriate enhancement of the light field amplitude given by
the factor $\sqrt\eta$ \cite{Moiseev2010}
($\langle\hat{E_2}\rangle=\sqrt\eta\langle\hat{E_1}\rangle$).
 By taking into account this requirement in Eqs. (\ref{eqn3AbsorbSystem2}),
(\ref{eqn4AbsorbSystem2}) and assuming
$G_{2,\mu}(\Delta_{\mu})=G_{2}(\Delta_{1})$,  we find five basic STR conditions
for the echo emission (written in the \emph{first} possible form):

\textit{i-c)} $\tilde\Delta_2\rightarrow-\eta\tilde\Delta_1$,

\textit{ii-c)} $Z \rightarrow -z$,

\textit{iii-c)} $\tau_2\rightarrow -\tau_1/\eta$,

\textit{iv-c)}
$\Omega_{2}(-\tau_2/\eta)/\Delta_{0,2}=\sqrt{\eta}\Omega_{1}(\tau_1)/\Delta_{0,1
}$.

\textit{v-c)} $\hat{E_{2,0}}\rightarrow - \sqrt{\eta}\hat{E_{1,o}}$.

 Also we can find \emph{second} form of STR transformation. Here,by saving the
first three conditions $\textit{i-c)},\textit{iii-c)}$, we rewrite two others
as follows:
$\textit{iv-c)}^{\prime}
\Omega_{2}(-\tau_2/\eta)/\Delta_{0,2}=-\sqrt{\eta}\Omega_{1}(\tau_1)/\Delta_{0,1
}$,
$ \textit{v-c)}^{\prime} \hat{E_{2,0}}\rightarrow  \sqrt{\eta}\hat{E_{1,o}}$.
 It is seen that STR indicates increasing of the reading laser field amplitudes
by the factor $\sqrt{\eta}$.

 Finally we write STR symmetry in a \emph{third} form. By saving the first four
conditions $\textit{i-c)}- \textit{iv-c)}$  we add two new conditions:
$\textit{v-c)}^{\prime\prime} \hat{E}_{2,o}\rightarrow =
\sqrt{\eta}\hat{E}_{1,o}$, $\textit{vi-c)}^{\prime\prime}
\hat{M}_{12,2}\rightarrow - \hat{M}_{12,1}$.
 Concerning the new STR conditions $\textit{i-c})-\textit{v-c})$, we note that
the fourth condition \textit{iv-c)} arises for preserving the interaction
strength of the echo field with atomic system characterized by new IB bandwidth
on this stage.
 It is worth noting the used initial relation between the bandwidths of IB
$\Delta_{1,in}$ and input light signal $\delta \omega_1$ is held for the atomic
and light parameters $\Delta_{2,in}$ and $\delta \omega_2$ at the echo emission
stage.
For example, the scaling spectral factor $\eta>1$ will decrease the interaction
strength of light with atoms due to larger IB bandwidth  $\Delta_{2,in}=\eta
\Delta_{1,in}$.
However increasing the Rabi frequince in accordance with $\textit{iv-c})$ is possible to compensate this negative effect for time reversibility. We change Rabi frequency of reading laser pulse by the factor $\sqrt\eta$. In this case we preserve the same absorption coefficient for both stages.
Thus preserving the condition $\textit{iv-c})$ (or its similar form
$\textit{iv-c)}^{\prime}$) via appropriate changing the Rabi frequency of the
control laser field $\Omega_{2}$ or optical detuning $\Delta_{0,2}$, we can
realize the unitary c/d with wavelength conversion of the signal light pulse.
 Similar consideration is  valid for the \emph{second} and \emph{third}
forms of STR  transformations based on the additional conditions:
$\textit{iv-c)}^{\prime}-\textit{v-c)}^{\prime}$, or
$\textit{v-c)}^{\prime\prime}-\textit{vi-c)}^{\prime\prime}$.

 The discussed new conditions clearly demonstrate considerable opportunities
for realization of STR dynamics in the studied scheme of PEQM.
The question arises about the possibility of such STR symmetry  for other
realizations of the Raman echo QM scheme.

\subsection{Longitudinal broadening}

 Following \cite{Moiseev2010,Hosseini2009}  in this case we can switch the
initial IB  $ G_{2,1}=\delta (\tilde\Delta_{1} -  \chi_{1} Z)$ to the new shape
with inverted and scaled spatial gradient $ G_{2,2}=\delta (\tilde\Delta_{1}
+\eta \chi_{1} Z)$ with compression factor $\eta$ ($\chi_{2}=-\eta\chi_{1}$ so
the light pulse compression occurs for $\eta >1$ and decompression is for $\eta
< 1$).
 By using IBs $ G_{2,1}$ and $ G_{2,2}$  in Eqs.
(\ref{eqn3AbsorbSystem2}),(\ref{eqn4AbsorbSystem2}) we find that all the four
basic considerations of STR $\textit{i-c})-\textit{v-c})$ with two supplement
conditions are met for this type of IB if echo field is irradiated in the
backward direction (see also Eq. (\ref{echo-2})).
Thus, the four conditions constitute the generalized time reversal of CRIB based
techniques and open a perfect method for quantum c/d of the single photon light
fields.

It is worth noting a possibility of QM realization using Raman atomic
interactions in the media with time controlled refractive index
\cite{Clark2012}.
So one can also realize the STR of signal pulse in this scheme by combining the
time modulation of refractive index with appropriate intensity variation of the
control laser field determined by the conditions $\textit{i-c})-\textit{v-c})$.
For efficient realization of the described schemes, we have to evaluate the
optimal physical parameters where the ideal system of Eqs.
(\ref{eqn3AbsorbSystem2}), (\ref{eqn4AbsorbSystem2}) can be used for real atomic
systems.

\section{Efficiency of quantum waveform conversion}

\subsection{The influence of inhomogeneous broadening on $1\leftrightarrow3$
transition }
The discussed ideal scheme of generalized STR assumes a negligibly small IB
width of the atomic  transition $1\leftrightarrow3$ and sufficiently large
optical spectral detuning $\Delta_{0,1}$.
Also it is important to optimize the switching rate of control laser fields in
order to provide an accessible quantum efficiency.
Below we perform such analysis following the original system of equations
(\ref{eqn2AbsorbSystem})-(\ref{eqn3AbsorbSystem}) and conditions
$\textit{i-c})-\textit{v-c})$ of the ideal pulse compression.

By using  the initial condition $\langle\hat R_{12}(-\infty)\rangle=\langle
R_{13}(-\infty)\rangle=0$ in Eqs.
(\ref{eqn2AbsorbSystem})-(\ref{eqn3AbsorbSystem}) with constant control  laser
field $\Omega_1(\tau)=\Omega_{1,0}$ and applying Fourier transformation $
\tilde{B}(\nu,Z) =  \int^{+\infty}_{-\infty}  B(\tau_1,Z) e^{i \nu \tau_1}
d\tau_1$ (where $\hat B(\tau,Z)$ is atomic or field operator and
$\tilde{B}(\nu,Z)=\langle \hat{\tilde{B}}(\nu,Z)\rangle $) we get the solution

\begin{eqnarray}
A_{1,0}(\tau_1,Z)  =  \nonumber \\
\frac{1}{2 \pi} \int\limits_{- \infty}^{\infty}
d \nu
 \tilde{A}_{1,0}(\nu,0)
 \exp\left(  -i \nu \tau_1 - \frac{1}{2}\int_{0}^{Z} \alpha_1 (\nu,Z)d Z
\right),
\label{absorbitionField} \\
R_{12}(\delta_1,\Delta_1;\tau_1,Z)  = \nonumber \\
\int\limits_{-
\infty}^{\infty} d \nu
 \frac {g \Omega_{1,0} \tilde{A}_{1,0}(\nu ,0)
 \exp \left(  -i \nu \tau_1 - \frac{1}{2}\int_{0}^{Z} \alpha_1 (\nu,Z)d Z
\right)   }
 { 2 \pi  \left\{ ( \Delta_{0,1} +\delta_1-\nu - i\gamma_{31} )(
\Delta_1-\nu-i\gamma_{21})
-\Omega^{2}_{1,o} \right\} }, \label{absorbitionCoherence_12}\\
R_{13}(\delta_1,\Delta_1;\tau_1,Z)  =
\nonumber \\
\int\limits_{- \infty}^{\infty}
d \nu
 \frac {g \left(\Delta_1-  \nu -i\gamma_{21} \right) \tilde{A}_{1,0}(\nu ,0)
 \exp\left(  -i \nu \tau_1 - \frac{1}{2}\int_{0}^{Z} \alpha_1(\nu,Z)d Z \right)
}
 { 2 \pi \{( \Delta_{0,1} +\delta_1-\nu - i\gamma_{31} )(
\Delta_1-\nu-i\gamma_{21}) -\Omega^{2}_{1,o}\}}
\label{absorbitionCoherence_13},
 \end{eqnarray}

\noindent
where for completeness we have added the phenomenological decay constants
$\gamma_{21}$ and $\gamma_{31}$ for the atomic coherences $R_{12}$, $R_{13}$ and
effective absorption coefficient on the Raman transition,

\begin{equation}
 \alpha_1( \nu,Z ) =-i\beta \iint \limits^{\infty}_{-\infty}
\frac{G(\delta_1,\Delta_1;Z) (\Delta_1 -\nu -i \gamma_{21}) d \delta_1 d
\Delta_1 } {  \left( \Delta_{0,1}+ \delta_1 -\nu  -i\gamma_{31}\right)
(\Delta_1-\nu -i\gamma_{21}) - \Omega^{2}_{1,o} } .
 \end{equation}

\noindent
Excited atomic coherences are given by general solutions
(\ref{absorbitionCoherence_12}), (\ref{absorbitionCoherence_13}) for time
$\tau_1\gg\delta t_s$ after complete disappearance of the signal light pulse
($A_{1,0}(\tau_1 >\tau_o,Z)=0$).
General analysis of the atomic coherences can be performed on Eqs,
(\ref{absorbitionField})-(\ref{absorbitionCoherence_13}) and on similar
equations for the echo emission stage.
Here, we take into account that highest quantum efficiency is possible only for
sufficiently weak  decoherence of Raman transition
$\gamma_{21}\ll\Delta_{1,in}$.
We also assume a sufficiently large spectral detuning on the optical transition
$\Delta_{0,1}>\delta\omega_1,\Omega_1,\delta_{1,in},\Delta_{1,in}$ in Eqs. (12)
and (13). After such calculation we get the following relations for  atomic
coherences $$ R_{12} (\tau_1)\cong i \zeta_{12}g \tilde A_{1,0}(\nu_1,0) e^{-i
\nu_1 \tau_1} \exp \left( - \frac{1}{2} \int_{0}^{Z} \alpha_1(\nu_1, Z) dZ
\right),$$
and $$R_{13}(\tau_1)\cong\zeta_{13} R_{12}(\tau_1),$$
\noindent
where $\zeta_{12}= \Omega_{1,0} / (\Delta_{0,1} + \delta_{1} - \Delta_{1} + 2
\frac{ \Omega^{2}_{1,o} }{\Delta_{0,1} + \delta_{1} - \Delta_{1}} )$ and
$\zeta_{13}= \Omega_{1,0} /(\Delta_{0,1}+\delta_1- \Delta_1)$ in accordance with
Eq.(\ref{largedetunr_13}), $ \nu_1 \cong \Delta_1 -
\frac{\Omega^{2}_{1,o}}{\Delta_{0,1} + \delta_1 - \Delta_1} - i \gamma_{eff} $,
$ \gamma_{eff}=\gamma_{21}+\gamma_{31}(\frac{\Omega_{1,0}^2}{\Delta_{0,1}})$.
More detailed analysis of $\zeta_{12}(\delta_1)$ and $\zeta_{13}(\delta_1)$
shows that influences of these factors on $\delta_1$ will not lead to essential
effects for sufficiently large optical detuning $\Delta_{0,1}$.
This property of the Raman photon echo quantum storage reveals a nonadiabatic feature of the light-atom interaction in contrast to the EIT based quantum memory.
In this case ($\Delta_{\mu,0}\gg\delta_{1,in},\Delta_{1,in}$), with large
accuracy we can take $\zeta_{12}\cong\zeta_{13}\cong\Omega_{1,0}/\Delta_{0,1}$
and to simplify the complex absorption coefficient $\alpha_1(\nu,Z)$ as follows

\begin{equation}
\alpha_1(\nu,Z) \cong 2\beta \frac {\Omega_{1,0}^2}{\Delta_{0,1}^2}
\int_{-\infty}^{\infty}d \Delta_1
\frac {G_2 (\Delta_1,Z)}{(\Delta_1-\nu-i\gamma_{eff})},
\end{equation}

\noindent
where both the real $\alpha_{abs,1} (\nu,Z)$  and  the imaginary  $\alpha_{dis,1} (\nu,Z)$
parts of the complex absorption coefficient can  play a significant role in the
 effecient quantum storage of broadband signal light fields \cite{Moiseev2012}.

In the next step we  evaluate the quantum storage for the conditions which
are close to the case described by Eqs. (\ref{eqn3AbsorbSystem2}),
(\ref{eqn4AbsorbSystem2}).
Here, we identify the influence of IB on the optical transition to the excited
atomic coherences via the frequency $Re(\nu_{1})
\cong\Delta_1(1-\frac{|\Omega_{1,0}|^2}{\Delta_{0,1}^2})-\frac{|\Omega_{1,0}|^2}
{\Delta_{0,1}}+\delta_{1,R}$, where
$\delta_{1,R}=\delta_1(\Omega_{1,0}/\Delta_{0,1})^2$ (see for comparison Eq.
(\ref{eqn1AbsorbSystem2})).
 One can see that additional spectral detuning  $\delta_{1,R}$ in $Re(\nu_{1})$
will lead to additional irreversible dephasing of the macroscopic atomic
coherences $ R_{12} (\tau) $ and $ R_{13} (\tau) $ due to averaging over the
exponential factor $<\exp\{-i\delta_{1,R}\tau\}>$.
Thus, taking into account  $\delta_{1,R}$ in the macroscopic atomic coherence we
find that the echo field will decay proportionally to the factors
\begin{eqnarray}
\Gamma_{G}  & = &
\exp \left( -\frac{1}{4}(\Omega_{1,0}/\Delta_{0,1})^4
(1+\eta^2)(\delta_{1,in}(
\tau_{echo}(\eta)-\tau_{st}))^2\right), \label{EffGauss} \\
\Gamma_{L}  & = & \exp \left( - \frac{1}{2} (\Omega_{1,0}/\Delta_{0,1})^2
(1+\eta)
\delta_{1,in}(\tau_{echo}(\eta)-\tau_{st})\right), \label{EffLorentz}
\end{eqnarray}

\noindent
for the Gaussian (G) and Lorentzian (L) shapes of IB line on transition
$1\leftrightarrow3$, where we have assumed
$(\Omega_{2,0}/\Delta_{0,2})=\sqrt{\eta}(\Omega_{1,0}/\Delta_{0,1})$,$\tau_{echo
}(\eta)=\frac{1+1/\eta}{2}\tau_{echo}(1)$ is the time moment of echo pulse
irradiation \cite{Moiseev2010}  and $\tau_{st}$ is the storage time on the
long-lived atomic levels $1$ and $2$ (see Fig.\ref{RamFig1}).

Then we deal with the control laser field switching to transfer  of the
excited atomic coherence on the long-lived levels 1 and 2.
 This problem has not been discussed in the previous studies off resonant Raman
echo QM, while we show that it considerably influence the quantum
efficiency.
It is worth noting here that similar problem has been studied early in the QM
based on the electromagnetically induced transparency \cite{Gorshkov2008}.

\subsection{Switching the control fields}

 We assume sufficiently large time of the interaction   when the input
light field ($\tau_0 \gg \delta t_s $) and excited  macroscopic atomic coherence
 disappear completely in the medium ($ \langle \hat{E}_1(\tau_0,Z) \rangle=0$).
 At the same time the resonant atoms are excited to the state characterized by
nonzero atomic coherences $R_{12}$ and $R_{13}$.
Obviously, the ideal quantum storage will occur for complete adiabatic transfer of the optical coherence $R_{13}$ on the long-lived levels 1 and 2. Below we evaluate the realistic conditions of the perfect transfer of the optical coherence  during the both stages.

 Taking into account the vanished input light field in equations
(\ref{eqn2AbsorbSystem} and \ref{eqn1AbsorbSystem}), we  evaluate the influence
of switching rate with exponential decay of the control laser field $\Omega_1
(\tau_1) = \Omega_{1,0} e^{-k (\tau_1-\tau_o)}$
(where $\tau_o$ is a moment of time when the switching off procedure starts, $k$
is a switching rate).
Here, we have the following atomic equations

\begin{eqnarray}
\frac{d R_{12}}{d\tau_1} & = &  -i(\Delta_{1}-i \gamma_{21}) R_{12} +
i\Omega_{1,0} e^{-k(\tau_1 -\tau_o)} R_{13} \label{firstInSystem}, \\
\frac{d R_{13}}{d\tau_1} & = & -i(\Delta_{0,1} +\delta_{1} -i \gamma_{31})R_{13}
+ i\Omega_{1,0} e^{-k(\tau_1 -\tau_o)} R_{12}. \label{secondInSystem}
\end{eqnarray}

 The atomic evolution of Eqs. (\ref{firstInSystem}) and (\ref{secondInSystem})
will not be accompanied by macroscopic atomic coherence and coherent
irradiation of the stored light field, respectively, due to the strong rephasing
of excited atoms during the switching procedure.
\begin{figure}
\includegraphics[keepaspectratio,width=0.6\textwidth]{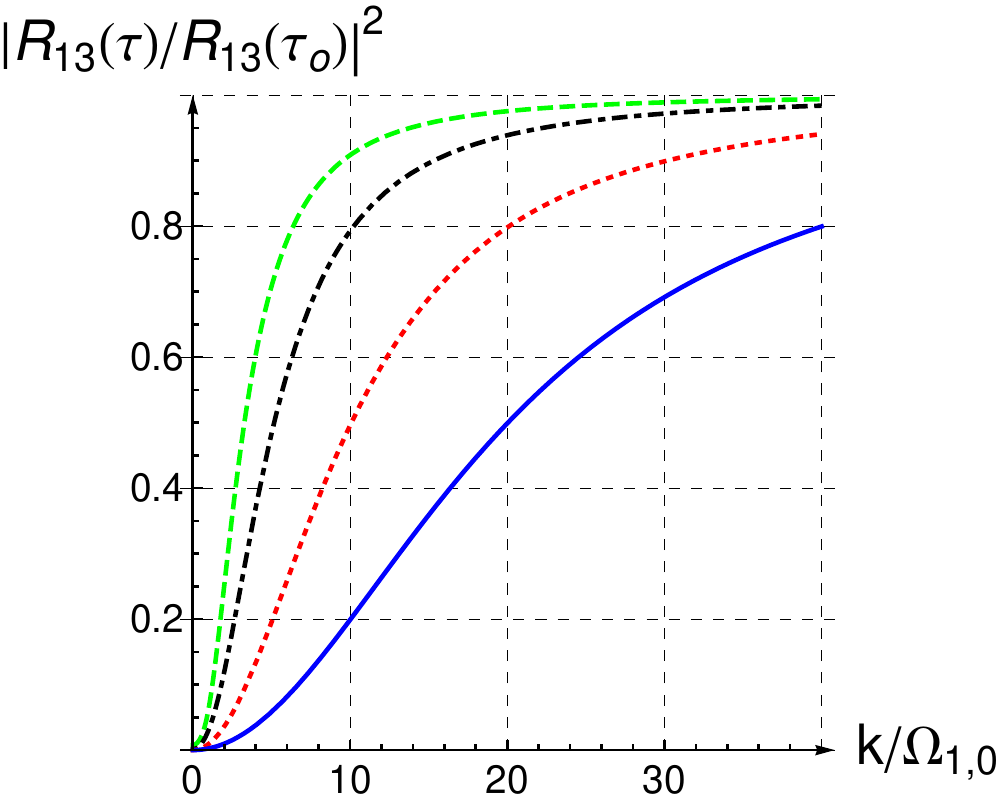}
\caption{ Atomic coherence $|R_{13}(\tau)/R_{13}(\tau_o)|^2$ after switching off
the first control laser field ($\tau-\tau_o\gg 1/k$) as a function of switching
rate in the units of Rabi frequency $k/ \Omega_1$. The coherence is given for
spectral detunings:  $\Delta_{0,1}/\Omega_{1,0}=3$ (green dashed line),
$\Delta_{0,1}/\Omega_{1,0}=5$
(black dot-dashed line), $\Delta_{0,1}/\Omega_{1,0}=10$ (red dotted line) and
$\Delta_{0,1}/\Omega_{1,0}=20$ (blue solid line).   }
\label{R13TEff}
\end{figure}
\begin{figure}
\includegraphics[keepaspectratio,width=0.6\textwidth]{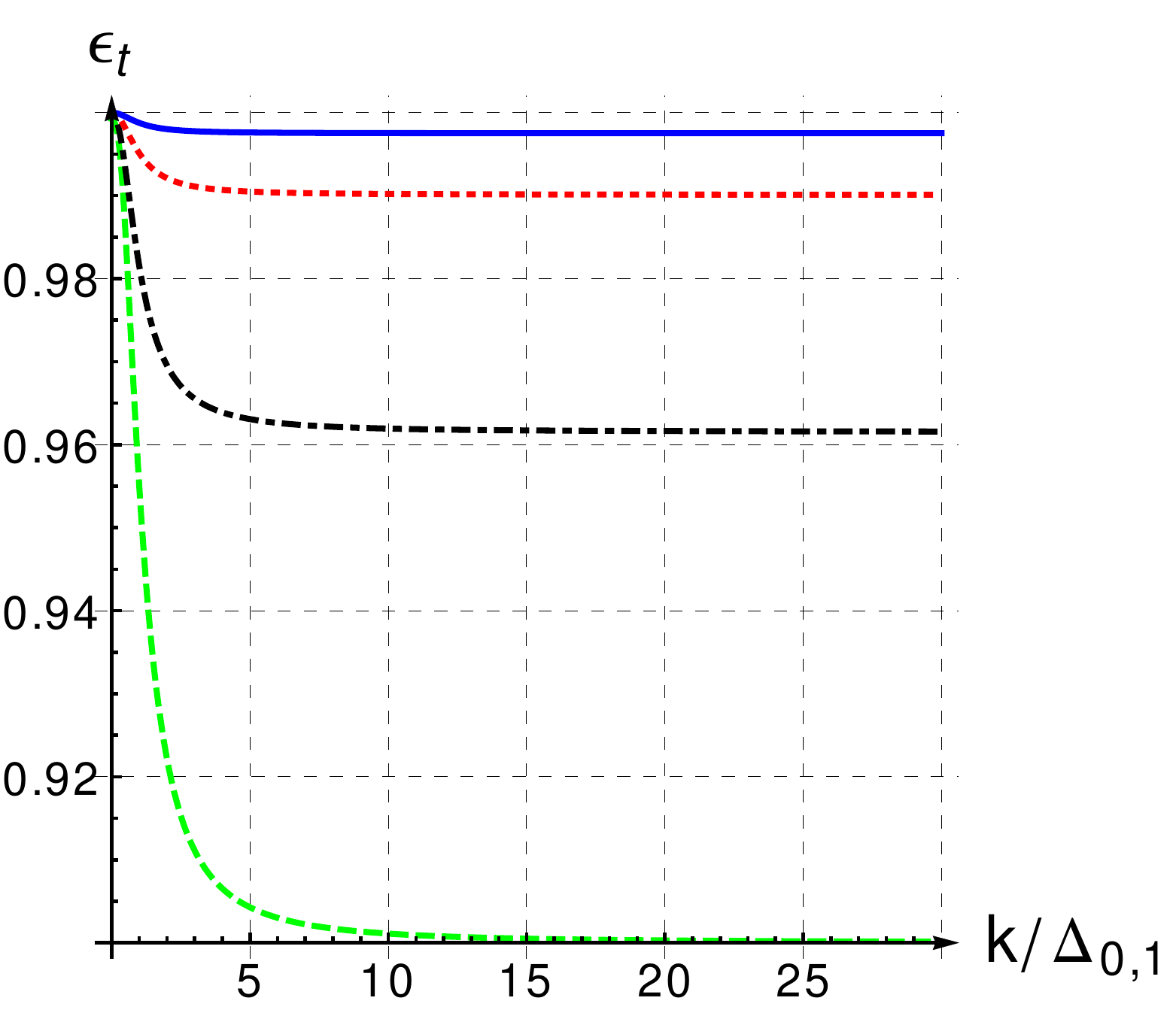}
\caption{ Transfer efficiency to the long-lived coherence after switching off
the first control laser field ($\epsilon_{t}= \vert R_{12}(\tau) \vert^2 /
\left( \vert R_{12}(\tau_0) \vert^2 +\vert R_{13 }(\tau_0) \vert^2 \right)$,
$\tau-\tau_o\gg 1/k$). The curves are given for switching rate in the units of
spectral detuning $k /\Delta_{0,1}$, $\Delta_{0,1}/\Omega_{1,0}=3$ (green
dashed line),
$\Delta_{0,1}/\Omega_{1,0}=5$
(black dot-dashed line), $\Delta_{0,1}/\Omega_{1,0}=10$ (red dotted line) and
$\Delta_{0,1}/\Omega_{1,0}=20$ (blue solid line).   }
\label{Teff}
\end{figure}
\noindent
After the substitutions $R_{12}= \tilde{R}_{12} e^{-(i\Delta_1+ \gamma_{21})
(\tau_1 - \tau_o) }$,
$R_{13}= \tilde{R}_{13} e^{-(i\Delta_1+ \gamma_{21}) ( \tau_1 - \tau_o )}$,  we
transfer to the new  atomic variables
$\tilde{R}_{12} = \chi^{(1+i\tilde\alpha)/2} Y(\chi)$, $\tilde {R}_{13} =
\chi^{(1+i\tilde\alpha)/2} \tilde{Y}(\chi)$ with appropriate timescale $ \chi =
e^{-k(\tau_1 -\tau_o)}\Omega_{1,0} /k$  (where $ \tilde\alpha \equiv  \left(
\Delta_{0,1}+\delta_{1}-\Delta_{1}-i(\gamma_{31}-\gamma_{21}) \right)/k $ ) that
leads Eqs. (\ref{firstInSystem}),(\ref{secondInSystem}) to the form of Bessel
equations

\begin{eqnarray}
\frac{d^2 Y}{d\chi^2} +\frac{1}{\chi} \frac{d Y }{d\chi} + \left(1-
\frac{(1+i \tilde\alpha)^2}{4 \chi^2}\right)Y =0, \\
\frac{d^2 \tilde{Y}}{d\chi^2} +\frac{1}{\chi} \frac{d \tilde{Y} }{d\chi} +
\left( 1- \frac{(-1+i \tilde\alpha)^2}{4 \chi^2} \right)\tilde{Y} =0.
\end{eqnarray}

\noindent
Using the well-known solutions  \cite{Bessel} of these equations,we find the
atomic coherences

\begin{eqnarray}
 R_{12}(\tau_1)  =  e^{ - \left( i\Delta_1+\gamma_{21} \right)(\tau_1 -\tau_0)}
\frac{(2k/\Omega_{1,0})^{ \frac{1 + i \tilde\alpha}{2}}}{ \Gamma(\frac{1 - i
\tilde\alpha}{2}) M} \nonumber \\
\left( J_{\frac{-1 + i \tilde\alpha}{2}}(\frac{\Omega_{1,0}}{k})
R_{12}(\tau_0)+iJ_{\frac{1 + i \tilde\alpha}{2}}(\frac{\Omega_{1,0}}{k})
R_{13}(\tau_0)   \right),
\end{eqnarray}

\begin{eqnarray}
 R_{13}(\tau_1)  =  i e^{ - \left( i(\Delta_{0,1}+\delta_1)+\gamma_{31}
\right)(\tau_1 -\tau_0)}
\frac{(2k/\Omega_{1,0})^{\frac{1 - i \tilde\alpha}{2}}}{ \Gamma(\frac{1 + i
\tilde\alpha}{2}) M} \nonumber \\
\left( J_{\frac{1 - i \tilde\alpha}{2}}(\frac{\Omega_{1,0}}{k})R_{12}(\tau_0)-i
J_{\frac{-1 - i \tilde\alpha}{2}}(\frac{\Omega_{1,0}}{k})R_{13}(\tau_0)
\right),
\end{eqnarray}
\noindent
where the complete switching off the control laser field occurs for
$\tau_1>\tilde\tau_1,
(\tilde\tau_1 -\tau_0)\gg 1/k$,
$J_{\frac{1 + i \tilde\alpha}{2}}(\frac{\Omega_{1,0}}{k})$ and  $\Gamma
\left(\frac{1 + i \tilde\alpha}{2}\right)$ are the Bessel and Gamma functions
\cite{Bessel}, $M = J_{\frac{1 + i \tilde\alpha}{2}}(\frac{\Omega_{1,0}}{k})
J_{\frac{1 - i \tilde\alpha}{2}}(\frac{\Omega_{1,0}}{k})+
J_{\frac{-1 + i \tilde\alpha}{2}}(\frac{\Omega_{1,0}}{k}) J_{\frac{-1 - i
\tilde\alpha}{2}}(\frac{\Omega_{1,0}}{k})$.

Fig.\ref{R13TEff} demonstrates a decrease of the optical coherence
$R_{13}(\tau_1=\tilde\tau_1)$ with enhancement of switching rate $k$ for various
atomic detunings $\Delta_{0,1}$ (for weak atomic decoherence
$\gamma_{21}(\tilde\tau_1-\tau_0)\ll1$,
$\gamma_{31}(\tilde\tau_1-\tau_0)\ll1$ and  negligibly small IBs
$\delta_{1,in},\Delta_{1,in}\ll\Delta_{0,1}$) after complete switching off the
control laser field ($\tilde\tau_1-\tau_0\gg 1/k$).
Effective depopulation of $R_{13}(\tilde\tau_1)$ occurs for large enough
detuning $\Delta_{0,1}$ where almost adiabatic transfer of the optical coherence
occurs to the long-lived levels.
 The transfer efficiency $\epsilon_{t}$ is given by $\epsilon_{t}= \vert
R_{12}(\tilde\tau_1) \vert^2 / \left( \vert R_{12}(\tau_0) \vert^2+\vert R_{13
}(\tau_0) \vert^2 \right)$ together with Eq. (6) and Eqs. (20),(21) for
sufficiently large optical detuning.
 Remnant coherence $R_{13}(\tau)$ will not contribute to the echo signal
emitting in the backward direction due to the miss phasematching. While it is
still possible to think about using the remnant optical coherence in the case of
fast control laser field manipulations for the forward echo emission.
Thus, the control field switching can lead to irreversible losses in quantum
transfer to the long-lived atomic levels.

We evaluate the transfer efficiency $\epsilon_{t}$ as a function of switching
rate $k$ (see Fig. \ref{Teff}) for various spectral detunings $\Delta_{0,1}$.
Here we assume a negligibly  weak influence of relatively  small IBs
($\Delta_{2,in},\delta_{1,in}\ll\Delta_{0,1}$) on $\epsilon_{t}$.
 This figure demonstrates a negative influence of fast switching procedure to
the efficiency $\epsilon_{t}$ that is considerably suppressed for large optical
spectral detuning  $\Delta_{0,1}\gg\Omega_{1,0}$.
However, the transfer efficiency $\epsilon_{t}\approx0.96$ can occur for
relatively small optical detuning $\Delta_{0,1}/ \Omega_{1,0} \approx 5$ and
fast switching rate.
The switching rate $k \geq 20$ leads to apparent negative impact for
intermediate  optical detuning $\Delta_{0,1}/ \Omega_{1,0} \leq 10$  where
$\epsilon_t \leq 0.99$.
As seen also in Fig.\ref{Teff},  retardation of the switching off procedure
increases the transfer efficiency.
However, using  low-speed switching rate $k$ is limited by the relaxation
processes of the optical atomic coherence.
 Moreover, the optical spectral detuning is limited since we have provide a
large enough optical depth of the atomic medium in accordance with $\alpha_1
\sim \beta \left( \frac{\Omega_{1,0}}{\Delta_{0,1}} \right)^2 $.
Thus, we have to use an optimal exchange for choice of the optimal switching
rate in order to get maximum quantum efficiency for really used parameters of
the light-atom interaction and IBs.

\begin{figure}
\includegraphics[keepaspectratio,width=0.5\textheight]{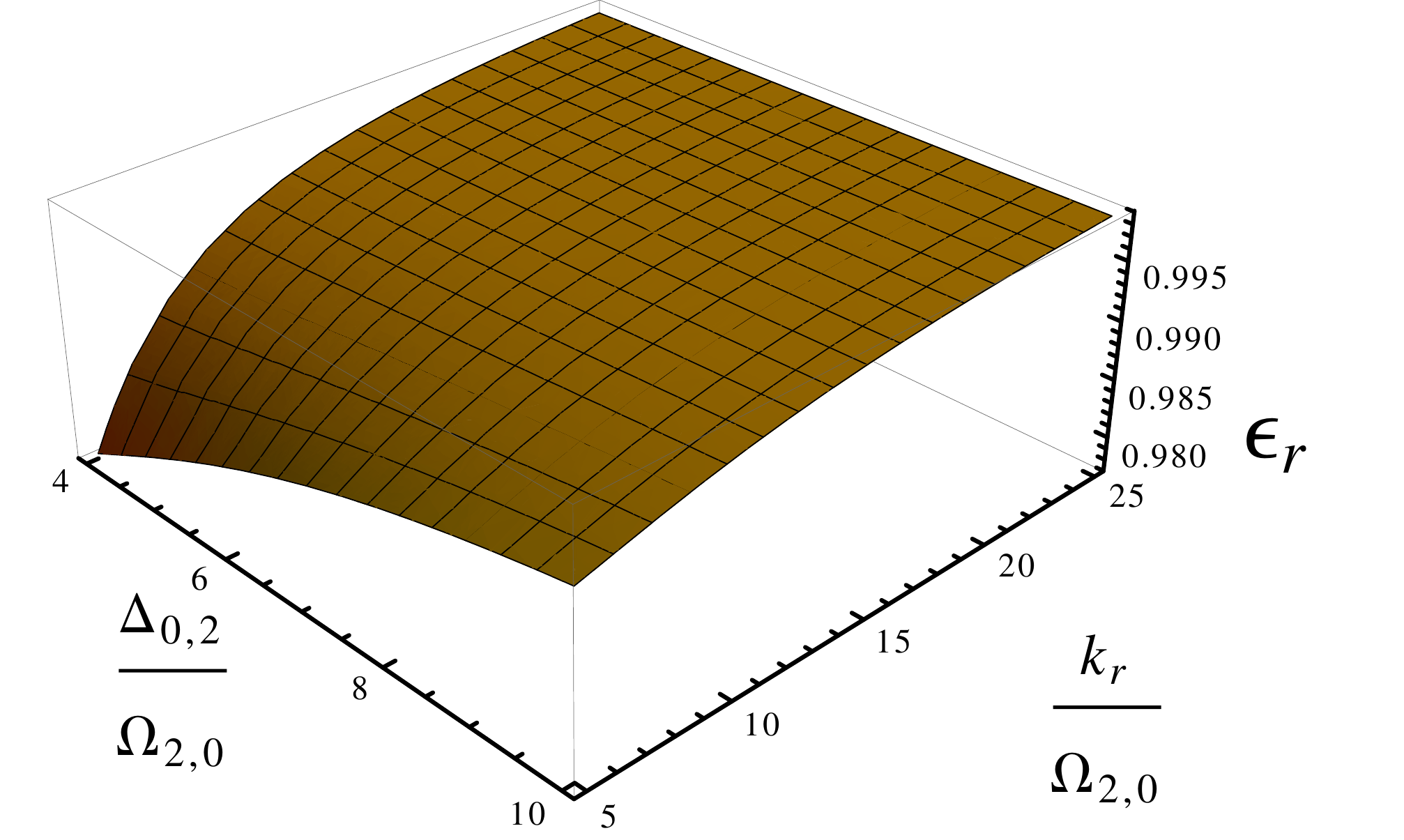}
\caption{  Quantum efficiency of switching on procedure $\epsilon_r$  as a
function of spectral detuning $\Delta_{0,2}/\Omega_{2,0}$ and   switching rate
$k_r/\Omega_{2,0}$}
\label{epsilonr_delta}
\end{figure}

\begin{figure}
\includegraphics[keepaspectratio,width=0.5\textheight]{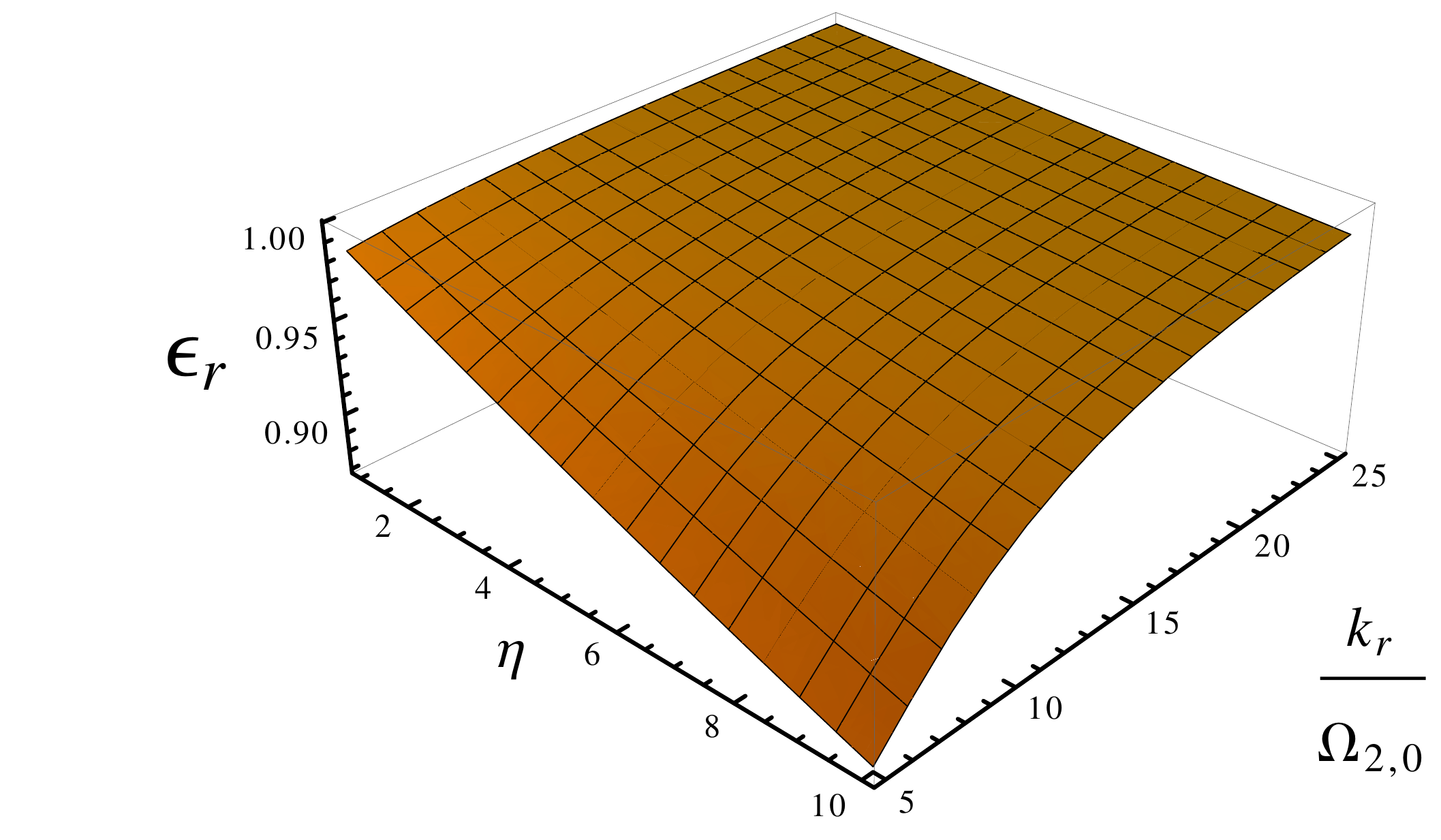}
\caption{  Quantum efficiency of switching on procedure $\epsilon_r$  as a
function of scaling  factor $\eta>1$  and   switching  rate $k_r/\Omega_{2,0}$
where  $\Delta_{2,o} = 6.5 \Omega_{1,0}$   }
\label{epsilonr_eta1}
\end{figure}

\begin{figure}
\includegraphics[keepaspectratio,width=0.5\textheight]{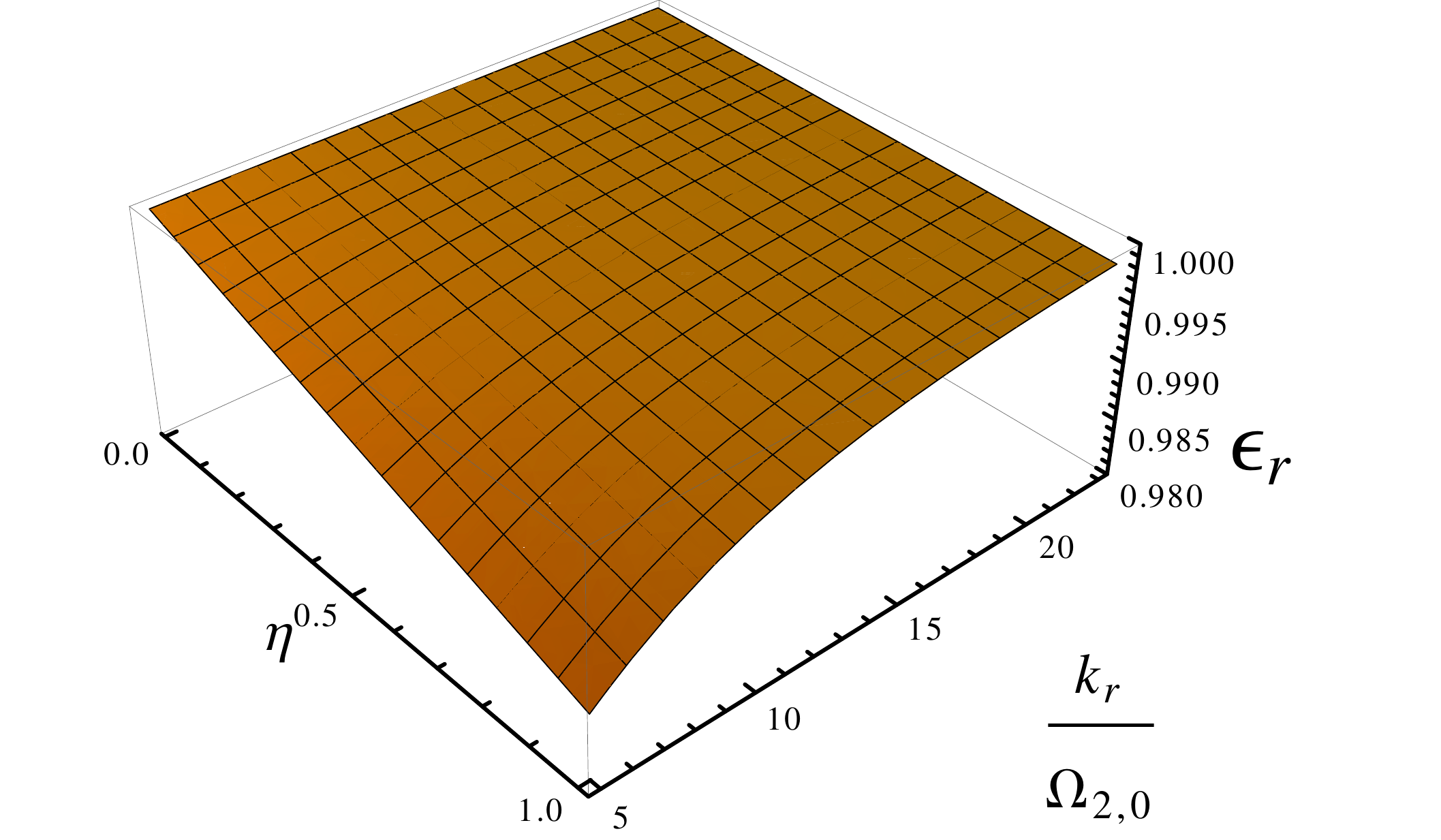}
\caption{  Quantum efficiency of switching on procedure $\epsilon_r$  as a
function of scaling  factor $\eta<1$  and   switching  rate $k_r/\Omega_{2,0}$
where  $\Delta_{2,o} = 6.5 \Omega_{1,0}$ }
\label{epsilonr_eta2}
\end{figure}

\begin{figure}
\includegraphics[keepaspectratio,width=0.4\textheight]{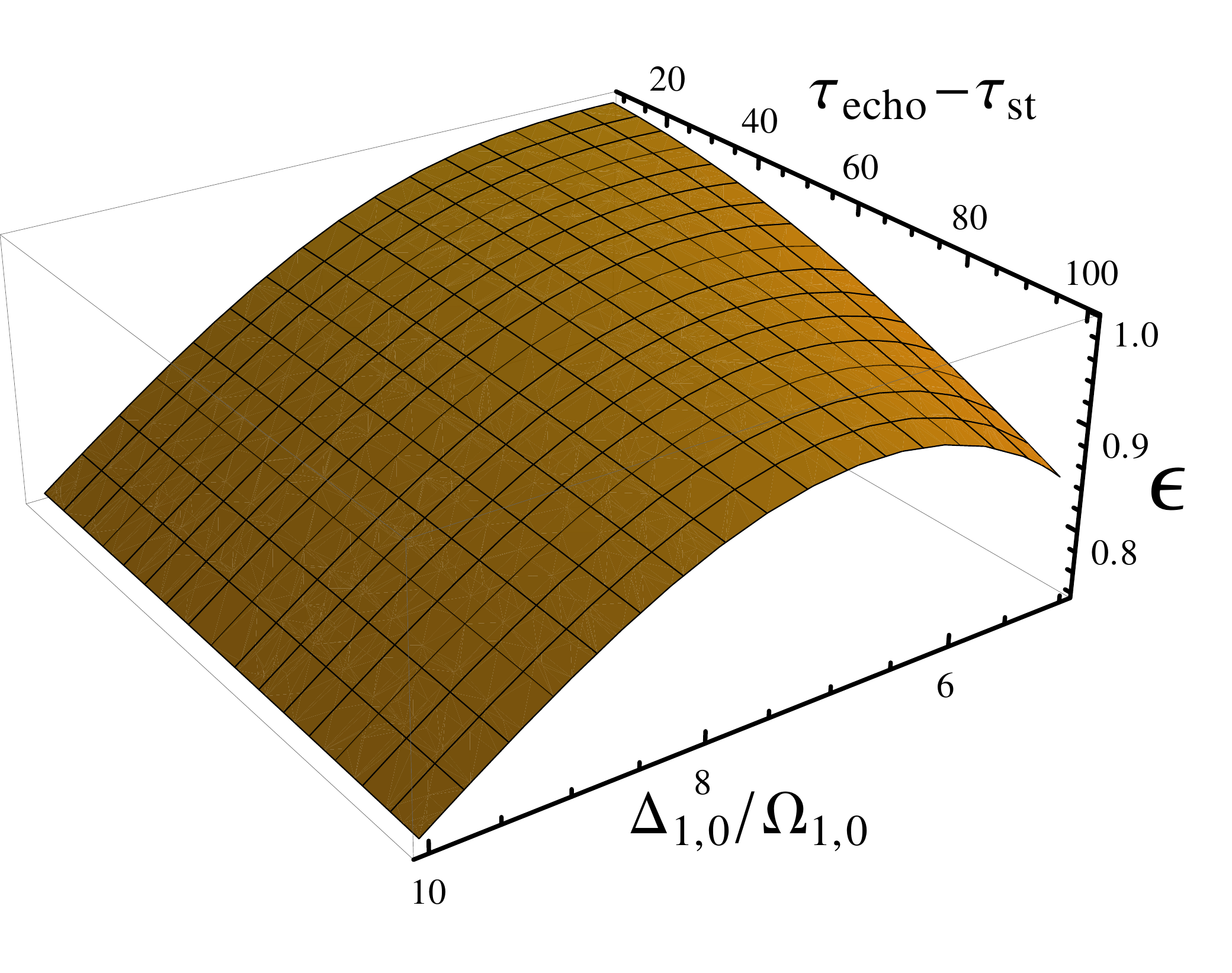}
\caption{ Quantum efficiency of the waveform conversion in scalable time
reversal as a function spectral detuning $\Delta_{0,1}/\Omega_{1,0}$ and
interaction time in the units $(\tau_{echo}-\tau_{st}) \Omega_{1,0}$ for
switching rate $k/\Omega_{1,0}=1$, optical depth $\tilde \kappa=200$, and Gaussian
IB shape with spectral width $\delta_{1,in}=0.1 \Omega_{1,0}$. }
\label{Queff1}
\end{figure}

\begin{figure}
\includegraphics[keepaspectratio,width=0.45\textheight]{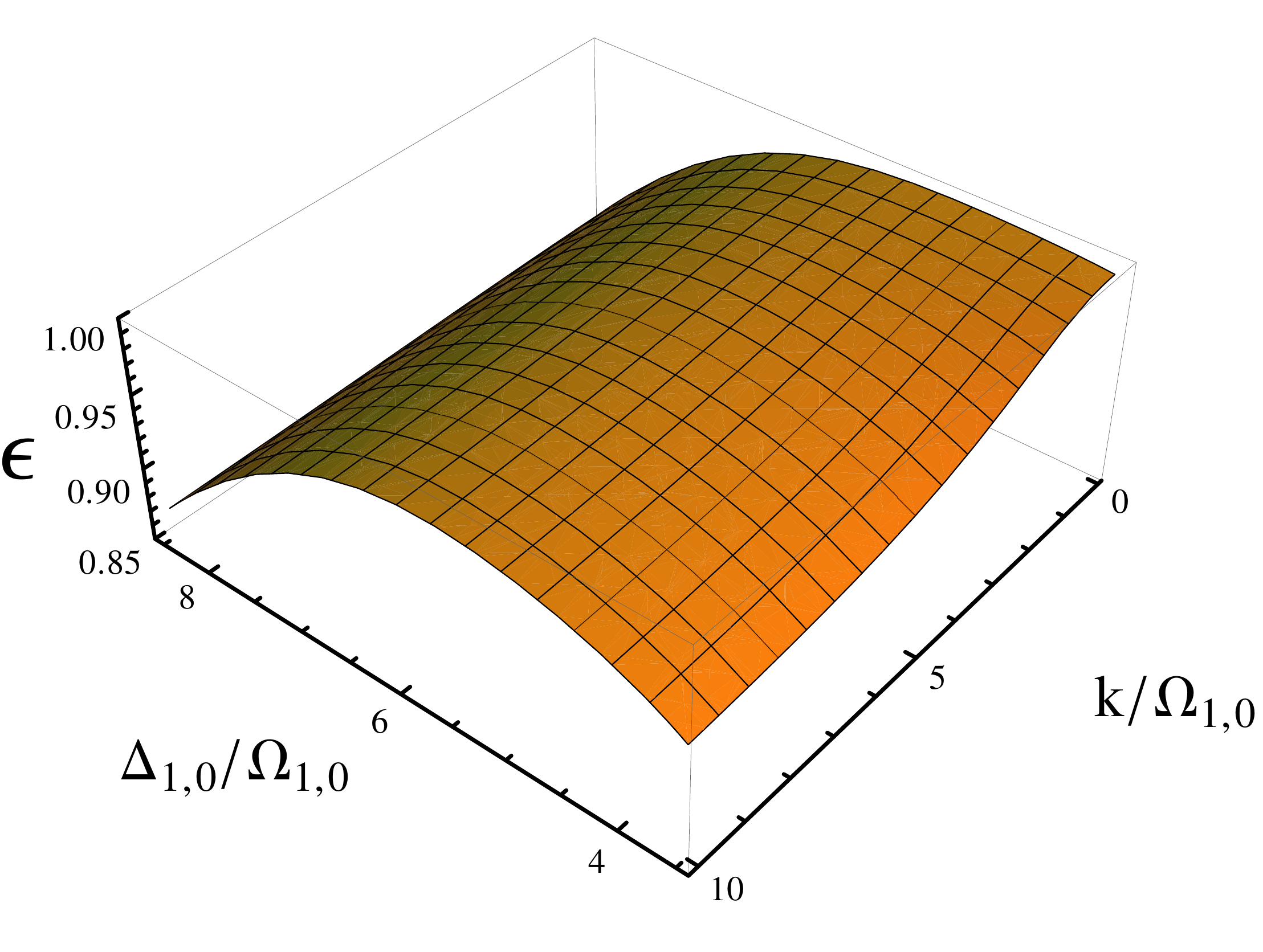}
\caption{ Quantum efficiency as a function of switching rate $k/\Omega_{1,0}$
and spectral detuning $\Delta_{0,1}/\Omega_{1,0}$ for short interaction  time
$(\tau_{echo}-\tau_{st}) \Omega_{1,0}=10$, optical depth $\tilde \kappa=200$, and
Gaussian IB shape with spectral width  $\delta_{1,in}=0.1 \Omega_{1,0}$. }
\label{Queff2}
\end{figure}

Let us assume  the control reading laser field is switched exponentially
$\Omega_2
(\tau_2) = \Omega_{2,0} e^{k_r (\tau_2-\tilde\tau_2)}$ with the switching rate
$k_r$. Evolution of atomic coherences are described by the atomic equations
which are quite similar to Eqs. (16)-(19) where $k$ is replaced by $"-k_r"$ and
$\Delta_1$ - by $"-\eta\Delta_1"$.
By taking into account initial coherence on the long-lived level $R_{12}
(\tilde\tau_1)$, we find  the solution for atomic equations  after  the
reading control field is switched on at the moment of time $\tau_2=\tilde\tau_2$

\begin{eqnarray}
R_{12} (\tilde\tau_2,\delta_1,\Delta_1,Z)  & = &  C_{1,2}  R_{12}
(\tilde\tau_1,\delta_1,\Delta_1,Z),
\\
R_{13} (\tilde\tau_2,\delta_1,\Delta_1,Z)  & = & i C_{1,3}  R_{12}
(\tilde\tau_1,\delta_1,\Delta_1,Z),
\end{eqnarray}

\noindent
where the functions $C_{1,2}$ and $C_{1,3}$ are
\begin{eqnarray}
C_{1,2}  & = &  \left( \frac{  \sqrt{\eta} \Omega_{1,0}}{2 k_r}
\right)^{\frac{1-i\Delta_{0,2}/k_r}{2}}
  \Gamma\left( \frac{1+i\Delta_{0,2}/k_r}{2} \right)
J_{\frac{i\Delta_{0,2}/k-1}{2}} \left(
\frac{  \sqrt{\eta} \Omega_{1,0}}{k_r}  \right),
\\
C_{1,3}  & = & i  \left( \frac{ \sqrt{\eta} \Omega_{1,0}}{2 k_r}
\right)^{\frac{1-i\Delta_{0,2}/k_r}{2}}
  \Gamma\left( \frac{1+i\Delta_{0,2}/k_r}{2} \right)
J_{\frac{i\Delta_{0,2}/k_r}{2}} \left(
  \frac{  \sqrt{\eta} \Omega_{1,0}}{k_r} \right),
\end{eqnarray}

\noindent
are given with accuracy of the same phase factor $e^{i\phi}$ and it was assumed
$\Omega_{2,0}=\sqrt{\eta} \Omega_{1,0}$. Here it is worth noting an important
property for the functions:
$|C_{1,2}|^2+|C_{1,3}|^2=1$.
Further evolution of atoms and light is accompanied by rephasing of the macroscopic coherences of the echo field emission. 
The optimal switching on procedure of the reading laser pulse can be clarified from general analysis of the echo emission.

\subsection{ Echo field emission}

Here we use the coupled system of light-atom equations with new atomic detuning
$\Delta_2=-\eta\Delta_1$ on the long-lived transition $1\leftrightarrow2$ and
echo field emission in the backward direction

 \begin{eqnarray}
 \frac{ \partial R_{13}}{ \partial  \tau_2} & = & - i  \left(  \Delta_{0,2}  +
\delta_1   \right) R_{13} + ig A_{2,0} + i  \Omega_{2,0} R_{12},
\\
 \frac{ \partial R_{12}}{ \partial  \tau_2} & = &  i   \eta \Delta_{1}
 R_{12}   + i    \Omega_{2,0} R_{13},
\\
\frac{\partial A_{2,0}}{ \partial Z} & =  & -i \frac{ \pi n g S}{ v_g}
\int \limits^{\infty}_{-\infty}  d \delta_1
d \Delta_{1} G( \delta_1,
\Delta_1, Z ) R_{13},
 \end{eqnarray}

\noindent
where the initial condition for atomic coherences and irradiated field  are
determined by
$R_{12} (\tilde\tau_2)$ and $R_{13} (\tilde\tau_2)$ and $A_{2,0}
(\tilde\tau_2,Z)=0$
at time $\tau_2=\tilde\tau_2$.

We solve the system equations in spectral Fourier representation of the atomic
coherences
$R_{12(13)}(\nu,...,Z)=\int_{-\infty}^{\infty} d \tau_2 R_{12(13)}(\tau_2,...,Z)
e^{i\nu \tau_2}$
and field amplitude
$\tilde{A}_{2,0}(\nu,Z)=\int_{-\infty}^{\infty} d \tau_2 A_{2,0}
(\tilde\tau_2,Z) e^{i\nu \tau_2}$. After performing a number of standard
calculations \cite{Moiseev2004,Moiseev2010}
we find following equation for the Fourier component of echo field amplitude
$\tilde{A}_{2,0}(\nu,Z)$:

\begin{eqnarray}
\frac{\partial \tilde{A}_{2,0}(\nu,Z)}{ \partial Z}  =   -\frac{\beta}{2g} \int
\limits^{\infty}_{-\infty}
  d \delta_1 d \Delta_{1} G( \delta_1,
\Delta_1, Z )
  \nonumber \\
\bigg( \frac{  R_{13} ( \tilde\tau_2,\delta_1,\Delta_1,Z)
\left( \nu +   \eta \Delta_1 \right) +  \Omega_{2,0} R_{12} (
\tilde\tau_2,\delta_1,\Delta_1,Z )  }
{ \left( \Delta_{0,2}  + \delta_1 -\nu    \right) \left( \nu -   \eta \Delta_1
\right) +   \Omega^{2}_{2,o} } + \nonumber \\
\frac{    ig  (\nu +   \eta \Delta_{1} ) \tilde{A}_{2,0}(\nu,Z) }
{ \left( \Delta_{0,2}  + \delta_1  -\nu       \right)  (\nu +   \eta \Delta_{1}
) +   \Omega^{2}_{2,o}  } \bigg).
\end{eqnarray}

In evaluation of this equation we use the early discussed approximations
determined by sufficiently large optical detuning $\Delta_{0,1}$. Then similarly
to the calculation of integrals in Eqs.
(\ref{absorbitionField})-(\ref{absorbitionCoherence_13}), we find the following
equation for the echo field amplitude

\begin{eqnarray}
\frac{\partial \tilde{E}_{2,0}(\nu,Z)}{ \partial
Z}=\frac{\eta^{\prime}}{2\eta}\bigg(
2\sqrt{\tilde\epsilon}\frac{\alpha_{abs,1}(\nu/\eta,Z)}{\sqrt{\eta^{\prime}}}
\tilde{E}_{1,o}(-\nu/\eta,0) e^{i\nu \tau_{echo}-\frac{1}{2}\int_{0}^{Z} dz
\alpha_1(-\nu/\eta,z)}
  \nonumber \\
 + \alpha_1(\nu/\eta,Z) \tilde{E}_{2,0}(\nu,Z) \bigg), \label{echoeq}
\end{eqnarray}

\noindent
where optical Stark shift have  been taken into account in the new field
amplitude
$E_{2,0}=A_{2,0} \exp{\{-i(\beta Z/2\Delta_{0,2})\}}$ and
$\Omega_{2,0}/\Delta_{0,2}=\sqrt{\eta^{\prime}}\Omega_{1,0}/\Delta_{0,1}$,
$\alpha_{abs,1}(\Delta_1/\eta,z)=2\pi\beta\frac{\Omega_{1,0}^2}{\Delta_{0,1}^2}
G_2 (\Delta_1/\eta,z)$ is an absorption coefficient,
$\sqrt{\tilde\epsilon}=\sqrt{\epsilon_t} (C_{1,2} + i \frac{\sqrt{\eta\prime}
\Omega_{1,0}}{\Delta_{0,1}} C_{1,3})\Gamma_{G,L}$.

 Solution of Eq.(\ref{echoeq}) can be written for arbitrary ratio
$\eta^{\prime}/\eta$.
However,  we discuss only the case  $\eta^{\prime}=\eta$. Taking into account
$$\alpha_1(-\nu/\eta,z)+ \alpha_1(\nu/\eta,z)=
 2\alpha_{abs,1}(\nu/\eta,z),$$
in Eq.(\ref{echoeq}), we get the solution

\begin{eqnarray}
\tilde{E}_{2,0}(\nu,Z)=
\sqrt{\tilde \epsilon}\frac{\tilde{E}_{1,0}(-\nu/\eta,0)}{\sqrt{\eta}}
e^{i\nu\tau_{echo} +\frac{1}{2}\int_{0}^{Z} dz \alpha_1(-\nu/\eta,z)}
\nonumber \\
\bigg(  e^{-\int_{0}^{Z} dz^\prime \alpha_{abs,1}(-\nu/\eta,z^\prime)}
 -e^{-\int_{0}^{L} dz^\prime \alpha_{abs,1}(-\nu/\eta,z)}\bigg).
 \label{echo}
\end{eqnarray}

\noindent
Eq.(\ref{echo}) is valid for arbitrary type of IB including the transverse and
longitudinal IBs. In the case of almost constant optical depth
($\int_{0}^{L} dz^\prime \alpha_{abs,1}(-\nu/\eta,z^\prime)\cong \tilde\kappa$)
within  the input light pulse spectrum,  we get for the irradiated echo field
($E_{2,0}(\tau_2,Z=0)=\frac{1}{2\pi}\int_{-\infty}^{\infty} d \nu
\tilde{E}_{2,0}(\nu,0) e^{-i\nu \tau_2} $):

\begin{eqnarray}
E_{2,0}(\tau_2,0)=\sqrt{\tilde \epsilon \eta}
E_{1,0}(-\frac{(\tau_2-\tau_{echo})}{\delta t_{e}},0)
e^{-\gamma_{21}\tau_{echo}} (1-e^{-\tilde\kappa}),
 \label{echo-2}
\end{eqnarray}

\noindent
where $\delta t_e=\delta t_1/\eta$ is the temporal duration of the echo
pulse, the exponential factor $e^{- \gamma_{21} \tau_{echo}}$ is determined by relaxation of long-lived atomic coherence during storage time.
The result confirms general consideration of Section III based on formal
using of STR symmetry.
Using Eq.(\ref{echo-2}) we obtain the quantum efficiency:

\begin{equation}
\epsilon= \epsilon_{t} \epsilon_{r}\Gamma_{G,L}^2
e^{-2\gamma_{21}\tau_{echo}}|1-\exp(-\tilde \kappa)|^2,
\label{overaleff}
\end{equation}

\noindent
where the latter two factor in $\epsilon$ (phase relaxation of the atomic
transition $1\leftrightarrow 2$ and finite optical depth $\tilde \kappa$) are
quite typical for CRIB in particular for the Raman scheme
\cite{Moiseev2011,Nunn2008,Gouet2009}.
The first three factors (result of this work) describe an influence of switching
procedure and IB on the optical transition.
Here, the quantum efficiency of switching on is determined by
\begin{equation}
\epsilon_{r} = \vert C_{1,2} \vert^2  + \vert
\frac{\sqrt{\eta}\Omega_{1,0}}{\Delta_{0,1}} C_{1,3} \vert^2.
\label{effon}
\end{equation}

\noindent
Quantity $\epsilon_{r}$ demonstrates maximum efficiency for fast switching on of
the control laser field in contrast to the switching off procedure.

We can conclude from Eq. (\ref{effon}), that only small part  ($\sim\vert\frac{\sqrt{\eta}\Omega_{1,0}}{\Delta_{0,1}} C_{1,3}\vert^2 \ll \vert C_{1,3}\vert^2$) of the excited optical coherence  evolves
adiabatically and rephases with long-lived atomic coherence.
Such behavior is accompanied by frozen dephasing of this optical atomic
coherence. While another leaving part of the atomic coherence will not participate in the echo emission that decreases the echo field retrieval.
Particular properties of the switching on efficiency  are demonstrated on  the
Figs.\ref{epsilonr_delta},\ref{epsilonr_eta1} and \ref{epsilonr_eta2}.
Here it is worth noting that the quantum efficiency $\epsilon_{r}$  is independent
of scaling factor $\eta$ for sufficiently fast switching rate.
In the Figs. \ref{Queff1},\ref{Queff2},  we evaluate overall efficiency only for
the fast rate of switching on.

Thus in the case of highly perfect STR, main influence to the quantum
efficiency
can be determined by the switching off procedure in accordance with the factor
$\sim\epsilon_{t}\epsilon_{r}$.
All other decoherent processes are almost repeated  for the echo emission
stage.
In particular as seen in Eqs. (16), (17), it is realized for the influence of
extra detuning $\delta_{1,R}$  with weak modification due to the scaling factor
$\eta$.

Efficiency of the quantum waveform conversion (\ref{overaleff}) is depicted in
the Fig. \ref{Queff1} using quite typical atomic parameters for experiments with
condensed and gaseous atomic systems.
Fig. \ref{Queff1} demonstrates possible optimal values of optical detuning
$\Delta_{0,1}$ and interaction time $\tau_{echo}-\tau_{st}$ in the presence of
the writing and reading control fields.
As seen in the figure, maximum efficiency takes place for relatively small
interaction time.
For larger interaction time, one can find a local maximum at the frequency
detuning $\Delta_{0,1}\approx 6.5 \Omega_{1,0}$.
This maximum arises due to suppression of the atomic dephasing caused by IB on
$1\leftrightarrow 3$, however further increasing of $\Delta_{0,1}$ reduces
considerably the effective optical depth.
For highly efficient waveform conversion (realized at small interaction time),
one can see large influence of the switching rate in Fig. \ref{Queff2} which
provides almost perfect  c/d and wavelength conversion of the signal light
pulse.

Leaving further discussion of STR symmetry we note that the quantum efficiency
$\epsilon$ characterizes  general features of perfect transformation for weak
quantum light pulse and can be used for studying the delicate interaction of
single and multi-pulse light fields with resonant media.
We note that relatively slow rate of the switching off procedure is preferable for 
storage stage since we need to transfer the excited coherence to the long-lived
levels.
However the read-out stage is needed in a fast switching rate.
Even the control laser field is turned off without presence of the signal light
field, the perfect storage requires an optimal switching of the writing laser field. The optimization problem of the control field shapes partically reminds the optical quantum
memory based on the electromagnetically induced transparency
[\cite{Gorshkov2008}].

\section{Conclusion}

Main idea of the proposed deterministic quantum waveform conversion is based on
the observed  \emph{scalable time reversal} (STR) symmetry of the light-atom
dynamics inherent to the absorption  of  input light field and subsequent echo
emission in the off resonant Raman echo QM.
It is worth noting the Maxwell's  equations reveal enhanced symmetries which have
been the subject of thorough long-term studies \cite{Fushchich1994}.
The observed STR symmetry is a new type of discrete symmetry for the light-atoms
dynamics which arises in the medium with specifically controlled spectral
properties of resonant atoms.
STR symmetry demonstrates various physical possibilities for its realization
that opens up a way for theoretically perfect compression/decompression of
signal
light pulses.
STR technique is naturally integrated with long-lived storage and can be
realized with additional manipulations such as wavelength conversion of the
light field.
The proposed STR technique can be also implemented with transformation of the
input photon wave packets into two-color single photon fields.
Multi-pulse readout of the stored light pulse can be also used with STR readout
that is quite  interesting for effective generation of time bin photonic qubits
in optical quantum communications.
Here, one can use reversible transfer of excited atomic coherence to an
additional long lived transition ($1\leftrightarrow4$) with subsequent readout
in the photon echo signals.
It is worth noting usefulness of these opportunities for quantum engineering
with broadband multi-photon fields \cite{Clausen2011,Saglamyurek2011}.
Finally note the discussed STR dynamics could open up new promising
possibilities for studying the  fundamental interactions of photons with atoms
due to possible using more intelligent experimental approaches.

\ack{
Authors would like to thank Russian Foundation for Basic Research
through grant no. 12-02-91700 for partial financial support of this work. 
}

\end{document}